\documentclass[aps, 10pt, prx, twocolumn, superscriptaddress,longbibliography, amsfonts,amssymb,amsmath,a4paper,floats,floatfix]{revtex4-2}

\usepackage[utf8]{inputenc}
\usepackage[english]{babel}
\usepackage{graphicx}
\usepackage{dcolumn}
\usepackage{bm}
\usepackage{siunitx}
\usepackage{xspace}
\usepackage{xcolor, color, colortbl}
\usepackage{braket}
\usepackage{booktabs}
\usepackage{notes2bib}
\definecolor{LightGray}{gray}{0.9}	
\usepackage{lineno}

\newcommand{\mos}{MoS\ensuremath{_{\mathrm{2}}}\xspace}
\newcommand{\vtg}{\ensuremath V_{\mathrm{TG}}}
\newcommand{\vbg}{\ensuremath V_{\mathrm{BG}}}

\DeclareSIUnit\angstrom{\text {\AA}}

\begin{document}

\title{
Gate-tunable band-edge in few-layer MoS$_2$}

\author{Michele Masseroni}%
    \email{These authors contributed equally to this work.}
    \affiliation{Solid State Physics Laboratory, ETH Z\"urich, 8093 Z\"urich, Switzerland}
\author{Isaac Soltero}%
    \email{These authors contributed equally to this work.}
    \affiliation{Department of Physics and Astronomy, University of Manchester, Oxford Road, Manchester, M13 9PL, United Kingdom}
    \affiliation{National Graphene Institute, University of Manchester, Booth St. E., Manchester, M13 9PL, United Kingdom}
\author{James G. McHugh}%
    \affiliation{Department of Physics and Astronomy, University of Manchester, Oxford Road, Manchester, M13 9PL, United Kingdom}
    \affiliation{National Graphene Institute, University of Manchester, Booth St. E., Manchester, M13 9PL, United Kingdom}
\author{Igor Rozhansky}%
    \affiliation{Department of Physics and Astronomy, University of Manchester, Oxford Road, Manchester, M13 9PL, United Kingdom}
    \affiliation{National Graphene Institute, University of Manchester, Booth St. E., Manchester, M13 9PL, United Kingdom}
\author{Xue Li}%
    \affiliation{Department of Physics and Astronomy, University of Manchester, Oxford Road, Manchester, M13 9PL, United Kingdom}
    \affiliation{National Graphene Institute, University of Manchester, Booth St. E., Manchester, M13 9PL, United Kingdom}
\author{Alexander Schmidhuber}%
    \affiliation{Solid State Physics Laboratory, ETH Z\"urich, 8093 Z\"urich, Switzerland}
\author{Markus Niese}%
    \affiliation{Solid State Physics Laboratory, ETH Z\"urich, 8093 Z\"urich, Switzerland}
\author{Takashi Taniguchi}
	\affiliation{Research Center for Materials Nanoarchitectonics, National Institute for Materials Science,  1-1 Namiki, Tsukuba 305-0044, Japan}
\author{Kenji Watanabe}
	\affiliation{Research Center for Electronic and Optical Materials, National Institute for Materials Science, 1-1 Namiki, Tsukuba 305-0044, Japan}
\author{Vladimir I. Fal'ko}%
    \affiliation{Department of Physics and Astronomy, University of Manchester, Oxford Road, Manchester, M13 9PL, United Kingdom}
    \affiliation{National Graphene Institute, University of Manchester, Booth St. E., Manchester, M13 9PL, United Kingdom}
\author{Thomas Ihn}
    \affiliation{Solid State Physics Laboratory, ETH Z\"urich, 8093 Z\"urich, Switzerland}
    \affiliation{Quantum Center, ETH Z\"urich, 8093 Z\"urich, Switzerland}
\author{Klaus Ensslin}
    \email{Contact author: ensslin@phys.ethz.ch}
    \affiliation{Solid State Physics Laboratory, ETH Z\"urich, 8093 Z\"urich, Switzerland}
    \affiliation{Quantum Center, ETH Z\"urich, 8093 Z\"urich, Switzerland}

\date{\today}

\begin{abstract}
Transition metal dichalcogenides (TMDs) have garnered significant research interest due to the variation in band-edge locations within the hexagonal Brillouin zone between single-layer and bulk configurations.
In monolayers, the conduction band minima are centered at the $K$-points, whereas in multilayers, they shift to the $Q$-points, midway between the $\Gamma$ and $K$ points. In this study, we conduct magnetotransport experiments to measure the occupation in the $Q$ and $K$ valleys in four-layer molybdenum disulfide (MoS$_2$).
We demonstrate electrostatic tunability of the conduction band edge by combining our experimental results with a hybrid $k\cdot p$ tight-binding model that accounts for interlayer screening effects in a self-consistent manner.
Furthermore, we extend our model to bilayer and trilayer MoS$_2$, reconciling prior experimental results and quantifying the tunable range of band edges in atomically thin TMDs.
\end{abstract}

\maketitle


\section{Introduction}

Transition metal dichalcogenides (TMDs) display unique electronic \cite{kim_high-mobility_2012, movva_high-mobility_2015, sebastian_benchmarking_2021} and optical properties \cite{wang_electronics_2012, geim_van_2013}. 
A key feature of TMDs is the presence of multiple valleys in their band structure, with the conduction band hosting valleys at both the $K$ points and the $Q$ points (located midway between the $\Gamma$ and $K$ points) of the first Brillouin zone, while the valence band exhibits valleys at the $K$ points and the $\Gamma$ point \cite{kormanyos_kp_theory_2015}. 
The relative energy levels of these valleys are strongly influenced by the number of layers, leading to layer-dependent modifications in the band structure \cite{mak_atomically_2010, splendiani_emerging_2010}.

In our previous studies, we systematically investigated the conduction band of single layer \cite{pisoni_interactions_2018}, bilayer \cite{pisoni_absence_2019}, and trilayer MoS$_2$ \cite{masseroni_electron_2021}. 
We found that electron transport predominantly occurred via the $K$ valleys of the conduction band, while the occupation of $Q$ valleys, predicted by theory  \cite{mattheiss_band_1973, li_electronic_2007, lebegue_electronic_2009, ellis_indirect_2011, yun_thickness_2012, sun_indirect--direct_2016}, was
not detected in our measurements. 
In these studies, we employed electrostatic gating to overcome the Schottky barrier that typically forms at the metal-MoS$_2$ interface. 
However, the influence of the gate-induced electric field on the band structure, particularly the relative energy shift of different valleys \cite{movva_tunable_2018}, was not considered.

Here, we investigate four-layer MoS$_2$, where we observe a gate-dependent band edge transition in the conduction band minima, shifting from the $Q$ valleys at low gate voltages to the $K$ valleys at higher voltages. 
This transition reveals the sensitivity of the conduction band to external electric fields, which we attribute to the differing atomic orbital compositions of the $Q$ and $K$ valleys, as well as interlayer screening effects. 
Our findings not only demonstrate the coexistence of both $Q$ and $K$ valleys in biased four-layer MoS$_2$, but also show how electrostatic gating can effectively tune the band structure in TMD-based devices. 
These results offer deeper insights into the electronic properties of multilayer MoS$_2$ and present potential avenues for valley-selective device applications.

\begin{figure}
    \centering
    \includegraphics{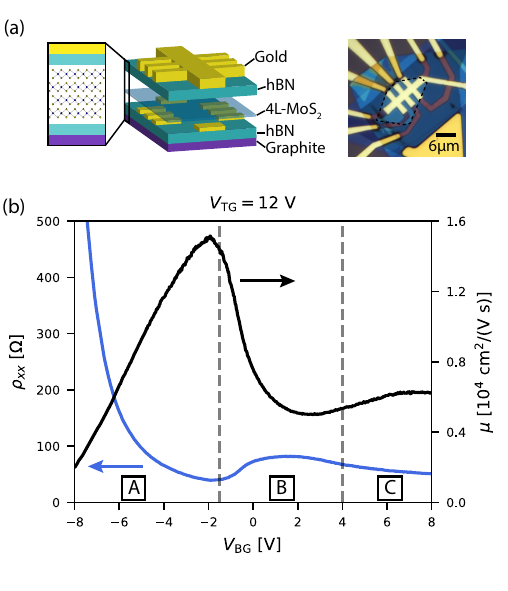}
    \caption{(a) Schematic view (left panel) and optical image of the sample (right panel). 
    In the right panel, the \mos layer is outlined with a black dashed line.
    The gate defines a conducting channel of width $W=\SI{2}{\micro m}$. The distance between adjacent contacts is $L_\mathrm{C-C}=\SI{3}{\micro m}$.
    (b) Four-terminal resistivity (blue) and mobility (black) as a function of the bottom gate voltage at $\vtg=\SI{12}{V}$. 
    The vertical dashed line mark the voltage at which additional bands are populated. This measurement was performed at a temperature of $T\approx\SI{100}{mK}$.
    }
    \label{fig:4L_Sample_TransferCharacteristic}
\end{figure}

\section{Results and discussion}

In this study, we use dual-gated, multi-terminal devices, as shown in Fig.~\ref{fig:4L_Sample_TransferCharacteristic}(a). The devices are heterostructures consisting of four-layer MoS$_2$ encapsulated in hexagonal boron nitride (hBN), all obtained by mechanical exfoliation. 
We conducted the experiments on two samples: Sample A, featuring a graphite bottom gate and a metallic top gate, and Sample B, with metallic gates on both sides. 
The results presented in the main text are from Sample A, which exhibits higher electron mobility and more pronounced Shubnikov-–de Haas oscillations (SdHO). 
Similar results obtained in Sample B are provided in the Supplementary Information.

One of the challenges of electronic transport experiments in MoS$_2$ devices is achieving ohmic contacts.
This issue is effectively addressed by employing a sample geometry with gated metallic contacts \cite{movva_high-mobility_2015, fallahazad_shubnikovhaas_2016, larentis_large_2018, pisoni_interactions_2018, masseroni_evidence_2023}. 
In such samples, finite conductance at cryogenic temperatures is achieved only at relatively large top gate voltages ($\vtg$), which are required to overcome the Schottky barrier at the metal-MoS$_2$ interface.

For this reason, in our experiments, we fix the top gate voltage and tune the density with the bottom gate voltage ($\vbg$), which does not affect the electron density in the region of the contacts due to the screening provided by the metallic contact pads.
This enables us to maintain low-resistive ($\sim\SI{1}{k\Omega}$) ohmic contacts for all densities in the Hall bar.
However, this sample geometry restricts the parameter space spanned by the gates.
As a consequence, the samples are typically operated under finite electric displacement field ($D$), which affects the electron distribution across layers in multilayer \mos, potentially affecting their electronic properties.

In Fig.~\ref{fig:4L_Sample_TransferCharacteristic}(b), we present the resistivity (blue) as a function of $\vbg$ measured at $\vtg = \SI{12}{V}$ and a temperature of $T\approx \SI{100}{mK}$. 
The resistivity reveals a non-monotonic dependence on $\vbg$, which is also evident in the transport mobility (black). 
The mobility increases roughly linearly in the voltage range $\vbg < \SI{-1.7}{V}$ (regime A in the figure), reaching a peak value of $\mu_\mathrm{peak} \approx \SI{1.5E4}{cm^2 , (V s)^{-1}}$ at an electron density of approximately $\SI{1E13}{cm^{-2}}$. 
Further increasing $\vbg$ results in a significant drop in mobility, which we attribute to the population of a second band, similar to our observation in three layer MoS$_2$ samples \cite{masseroni_electron_2021}.

\begin{figure}
    \centering
    \includegraphics[width=\columnwidth]{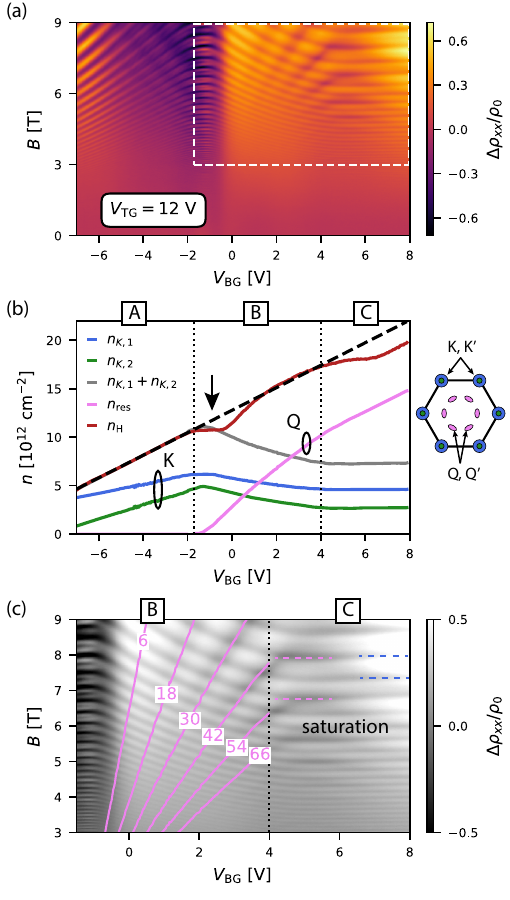}
    \caption{(a) $\Delta \rho_{xx}/\rho_0=[\rho_{xx}(B)-\rho_{xx}(0)]/\rho_{xx}(0)$ plotted as a function of $\vbg$ and $B$ at $\vtg=\SI{12}{V}$ and $T\approx\SI{100}{mK}$. 
    The white dashed frame highlight the zoom-in displayed in panel(c).
    (b) Electron densities as a function of $\vbg$.  The total density ($n_\mathrm{tot}$) is shown by the black dashed line, Hall density ($n_\mathrm{H}$) is represented by the red solid line, and the densities obtained from the SdHO frequencies ($n_{K,i}=2ef_{K,i}/h$ with $i=1,2$) by the blue and green solid lines. 
    The residual density ($n_\mathrm{res}=n_\mathrm{tot}-n_{K,1}-n_{K,2}$) is shown by the pink solid line.
    The schematic on the right shows the hexagonal Brillouin zone of \mos, with the electron in the $K$ and $Q$ valleys depicted by the colored pockets.
    (c) Zoom-in of panel (a), showing the Landau fan (pink solid lines) expected for the residual density, accounting for a degeneracy of $g=12$. 
    The dashed horizontal lines in regime C highlight the saturation of the densities in the $K$ and $Q$ valleys.
    }
    \label{fig:4L_Magnetoresistance_100mK}
\end{figure}

To investigate the contribution of the different bands in four-layer MoS$_2$, we examine the data obtained at finite magnetic fields.
In Fig.~\ref{fig:4L_Magnetoresistance_100mK}(a), we present the magnetoresistance $\Delta \rho_{xx}/\rho_0 = [\rho_{xx}(B) - \rho_{xx}(0)]/\rho_{xx}(0)$ as a function of $\vbg$ measured at $\vtg=\SI{12}{V}$. 
This measurement reveals an intricate pattern that originates from overlapping Landau fans.
The observation of multiple Landau fans confirm the existence of multiple electronic bands.

Each of those bands has a corresponding electron density which determines the frequency of the SdHO. We extract the densities by computing the Fast Fourier Transform (FFT) of $\Delta \rho_{xx}(B^{-1})$ (see details in the Supplementary Information) and plot the resulting densities in Fig.~\ref{fig:4L_Magnetoresistance_100mK}(b).
In the voltage range $\vbg<-\SI{1.7}{V}$ (regime A), the Fourier analysis reveals two electron densities, $n_{K,1}$ and $n_{K,2}$ (blue and green curves), both with a similar gate voltage dependence.
The densities are calculated using the relationship $n_{i} = g_{i} e f_i/h$, where $f_i$ is the frequency, $g_{i}$ is the degeneracy of band $i$, $e$ is the elementary charge and $h$ is Plank's constant. 
By comparing the sum $n_{K,1}+n_{K,2}$ (gray curve) to the total density $n_\mathrm{H}$ obtained from the Hall effect measurements (red curve), we determine a degeneracy of $g=2$ for both bands.
This degeneracy suggests that the bands are centered at the $K$ points of the Brillouin zone.

The density difference $n_{K,1}-n_{K,2}$ arises from the spin-orbit splitting of the $K$ valley.
By linear extrapolation, we estimate $n_\mathrm{K,1}\approx\SI{3.2E12}{cm^{-2}}$ at the onset of the density $n_{K,2}$, which aligns with the density required to populate the upper spin-orbit split band at the $K$ points observed in previous studies in monolayer \cite{pisoni_interactions_2018}, bilayer \cite{pisoni_absence_2019}, and three-layer \cite{masseroni_electron_2021} MoS$_2$. 
This further supports our interpretation.

Now, we turn our attention to regime B in Fig.\ref{fig:4L_Magnetoresistance_100mK}(b), where an additional band starts getting populated. This is evident in three distinct aspects. The first is the emergence of another Landau fan in Fig.\ref{fig:4L_Magnetoresistance_100mK}(a) (highlighted by the white dashed frame), with its origin around $\vbg\approx\SI{-1.7}{V}$. At the same gate voltage, the Hall density exhibits a plateau (marked by an arrow), which we attribute to the filling of defect states at the bottom of the newly occupied band. This results in the localization of electrons at the defects, preventing them from contributing to the Hall effect (see also the Supplementary Information for further discussion). Thirdly, the sum $n_{K,1} + n_{K,2}$  is no longer equal to the Hall density implying the presence of an extra electron pocket. 

We determine the degeneracy of the newly occupied band by calculating the residual density using the relation $n_\mathrm{res}=n_\mathrm{tot}-n_{K,1}-n_{K,2}$ [depicted by the pink curve in Fig.~\ref{fig:4L_Magnetoresistance_100mK}(b)] and using it to predict the Landau fan according to the equation
\begin{equation}
B_m(\vbg) = \frac{hn_\mathrm{res}(\vbg)}{e\, g\, m},
\end{equation}
where $m$ is an integer representing the Landau level index, and $g$ is the band degeneracy.
In Fig.~\ref{fig:4L_Magnetoresistance_100mK}(c), we plot the calculated Landau fan, where we match the minima of $\Delta \rho_{xx}/\rho_0$ by assuming $g=12$.
This large degeneracy suggests that the residual electron density originates from the $Q$ valley, which has a sixfold valley degeneracy.
To achieve a total degeneracy of 12, the bands must be spin degenerate, implying inversion symmetry.
However, while inversion symmetry is preserved in unbiased \mos samples with an even number of layers, the large displacement field in regime B ($D\sim\SI{1}{V/nm}$) is expected to break this symmetry.

To gain further insight, we implement a model to analyze the layer-resolved valley densities in four-layer \mos under various displacement fields and total electron densities.
This model incorporates self-consistent screening effects produced by charge accumulation in each layer, allowing us to capture the impact of inter-layer screening on the electron distribution. 
We construct hybrid $k \cdot p$ tight-binding Hamiltonians to simulate the electronic structure of the $K$ and $Q$ valleys.
These Hamiltonians were initially parametrized using density functional theory (DFT) calculations (see Supplementary Information) and included an on-site term to account for electrostatic band bending due to external fields.
Notably, DFT results confirmed the absence of hybridization between the $K$ valleys of different layers, consistent with prior results \cite{pisoni_absence_2019}. According to the findings in Ref. \cite{PhysRevB.110.L161404}, the spin-orbit splitting between $K$ bands, $\Delta_{\rm SO}^{K}$, and the monolayer energy offset between $K$ and $Q$ valleys, $E_{KQ}$, undergoes significant renormalization due to electron-electron exchange interactions. In line with this framework, we considered both of these quantities as layer-density dependent (see Supplementary Information).

The predictive accuracy of the model is based on two essential parameters: $t_0$, the inter-layer tunnel coupling for $Q$ valleys, and $E_{KQ}$. 
However, DFT-derived parameters exhibit a pronounced sensitivity to subtle variations in the \mos interlayer spacing and lattice constants.
This sensitivity introduces considerable uncertainty in their precise values, and, in fact, theoretical values alone did not align with our experimental results. 
To overcome this limitation, we varied $t_0$ and $E_{KQ}$ semi-empirically to achieve a better fit to the observed layer and valley densities.
Remarkably, we found that the distribution of electron densities across layers and valleys is highly sensitive to variations in $t_0$ and $E_{KQ}$.
This characteristic suggests a novel experimental approach for determining inter-layer tunnel coupling, which may also be extended to other TMDs.
The optimized parameters for four-layer \mos determined through this process are $t_0=\SI{0.16}{eV}$ and $E_{KQ}=\SI{0.21}{eV}$. 

\begin{figure}
    \centering
    \includegraphics{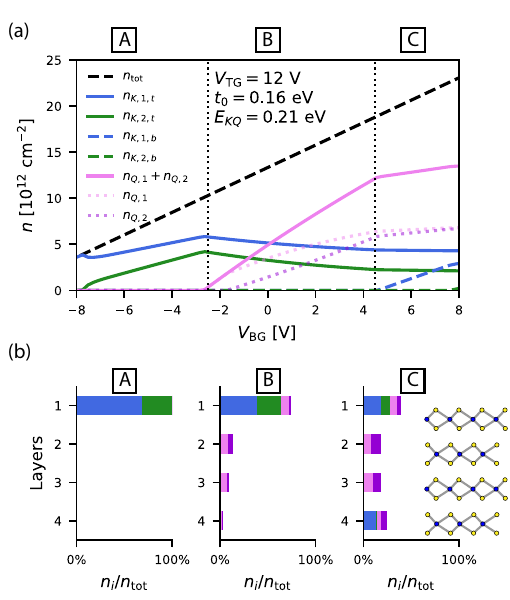}
    \caption{(a) Band densities obtained from the theoretical model using the parameters $E_{QK}=\SI{0.21}{eV}$ and $t_0=\SI{0.16}{eV}$. The densities are plotted as a function of $\vbg$ at $\vtg=\SI{12}{V}$, enabling a direct comparison with Fig.~\ref{fig:4L_Magnetoresistance_100mK}(b) . The densities in the $K$ valleys of the top layer (layer 1), $n_{K,1,t}$ and $n_{K,2, t}$, are shown as solid blue and green lines, respectively. The $K$ valley densities in the bottom layer (layer 4) are depicted with dashed lines following the same color code. The total density in the $Q$ valleys ($n_{Q,1}+n_{Q,2}$ for a direct comparison with the experiment) is represented by the pink solid line, while the individual components are shown by the dotted lines.
    (b) The three panels show the percentage of the total density distributed across the four \mos layers with the different bands encoded by the colors  [consistent with panel (a)].
    As a representation of the three regimes we selected specific gate voltage configurations: $(\vtg=\SI{12}{V}, \vbg=-\SI{6}{V})$ for regime A, $(\vtg=\SI{12}{V}, \vbg=\SI{0}{V})$ regime B, and $(\vtg=\SI{12}{V}, \vbg=\SI{8}{V})$ for 
    regime C.}
    \label{fig:Theory_Layer_Dist}
\end{figure}

The band densities obtained from our model are shown in Fig.~\ref{fig:Theory_Layer_Dist}(a) as a function of $\vbg$ for $\vtg=\SI{12}{V}$, allowing a direct comparison with the experimental densities in Fig.~\ref{fig:4L_Magnetoresistance_100mK}(b).
The model identifies three distinct operating regimes controlled by the bottom gate voltage, reflecting the experimental observations.
In regime A, where the voltage is applied asymmetrically between the top and bottom gates, the density is concentrated entirely in the $K$ valleys of the top layer [see the layer polarization shown in Fig.~\ref{fig:Theory_Layer_Dist}(b)]. 
In regime B, where the bottom gate voltage is close to zero, the $Q$ valleys begin to fill with electrons, and charge is transferred from the $K$ valley to the newly occupied band. 
Since the $Q$ valley states are hybridized between layers, the density is distributed across layers, favoring layers with lower potential energy.
In regime C, the $K$ valleys in the bottom layers are filled, leading to the saturation of the $K$ (top layer) and $Q$ valley densities, as observed in the experiment.

This model successfully captures the gate voltage dependence of the valley-specific densities and provide insights into the layer distribution of electron densities. 
In regime B, the model predicts an asymmetric density distribution across the layers, which results in a spin degeneracy lifting of $Q$ valley states when $D\neq 0$, due to SOC [see dotted lines in Fig.~\ref{fig:Theory_Layer_Dist}(a)].
This prediction aligns with a detailed analysis of the FFT spectrum (see Supplementary Information), where we observe a splitting of the SdHO frequency corresponding to the residual density. 

The experimentally observed splitting of the $Q$ valley is notably less pronounced than predicted by our model. 
This discrepancy may arise from crystal field effects at the interface with the hBN dielectrics \cite{rickhaus_gap_2019}.
In the Supplementary Information, we account for the influence of hBN by introducing an on-site potential energy difference, $\Delta_\mathrm{hBN}$, between the outer layers (which interface directly with hBN) and the inner layers. 
Incorporating this offset reduces the hybridization between the Q valleys of the outer and inner layers.
This adjustment results in a more symmetric layer distribution of the $Q$ valley electrons, thereby mitigating inversion symmetry breaking induced by the displacement field and leading to a weaker spin-orbit splitting.
Consequently, the model's prediction of spin-orbit splitting becomes more consistent with the experimental observations.

\begin{figure}
    \centering
    \includegraphics[width=\linewidth]{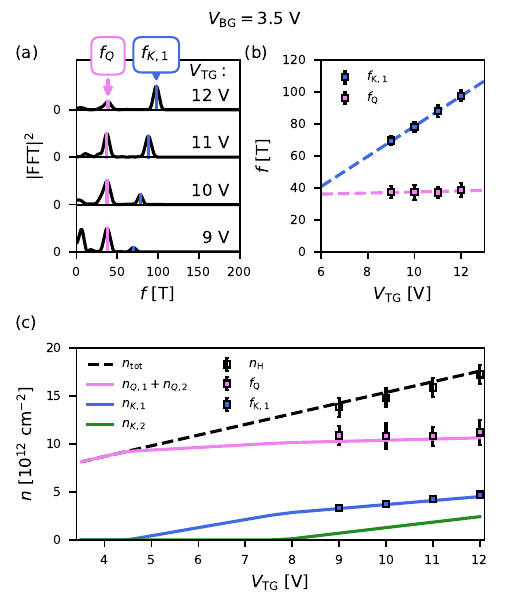}
    \caption{(a) FFT of $\Delta \rho_{xx}(B^{-1})$ at $\vbg=\SI{3.5}{V}$ and various $\vtg$. The peaks associated to the $Q$ and $K_{1}$ valleys are highlighted by the vertical pink and blue line, respectively. (b) The frequencies obtained from panel (a) plotted as a function of $\vtg$.
    (c) Comparison between the densities obtained from the model (solid lines) and the densities obtained from the experiment (square markers). The model parameters are the same of Fig.~\ref{fig:Theory_Layer_Dist}}
    \label{fig:TopGate_dependence}
\end{figure}

Next, we examine the effect of the top gate voltage, focusing on the densities in the $Q$ and $K$ valley of the top layer.
To track the top gate dependence of these densities, we present the FFT of $\Delta \rho_{xx}(B^{-1})$ at ${\vbg=\SI{3.5}{V}}$ for various $\vtg$ in Fig.~\ref{fig:TopGate_dependence}(a).
In this regime, the finite $\vbg$ allows us to observe oscillations originating from both $Q$ and $K$ valley states of the top layer without inducing charges in the $K$ valleys of the bottom layer. 
Figure~\ref{fig:TopGate_dependence}(b) shows that the $K$ valley frequency of the top layer increases with $\vtg$, while the $Q$ valley frequency is nearly constant due to interlayer screening.
These data suggest that the top gate voltage induces a transition of the band edge from the $Q$ valleys at low gate voltages to the $K$ valleys at higher voltages.

In Fig.~\ref{fig:TopGate_dependence}(c), we compare valley densities predicted by the model (solid lines) with experimental densities (square markers), showing a good agreement.  
Due to high contact resistance, the low $\vtg$ regime is experimentally inaccessible, so we rely on model predictions for this range.
According to the model, at low bias ($\vtg<\SI{4.5}{V}$), the $Q$ valley remains the only occupied band, confirming that the conduction band minimum in low-biased four-layer \mos is located at the $Q$ point and demonstrating the presence of a gate-induced $Q-K$ band edge transition.

Having established the relevance of the gate bias in determining the relative band alignment in multilayer \mos, we extended our model to bilayer and three-layer systems (see Supplementary Information) to address discrepancies between experiments and DFT-calculated band structures. 
Using consistent model parameters, $E_\mathrm{KQ}=\SI{0.21}{eV}$ and $t_0=\SI{0.16}{eV}$, we successfully reproduce the experimental findings across multiple layers.

For biased bilayer \mos, the model predicts that the $K$ valleys in top and bottom layers are occupied bands, while the $Q$ valleys are not, as observed in Ref.~\cite{pisoni_absence_2019}.
In our earlier study on three-layer \mos \cite{masseroni_electron_2021}, we observed electron transport predominantly through the $K$ valleys in the outer layers, although the band contributing to the electrons in the middle layer remained unidentified.
Our model reveals the missing element in our previous work, showing that both $K$ and $Q$ valleys were populated, as observed in the four-layer case.

\section{Conclusion}

In conclusion, our study provides a comprehensive understanding of the gate-tunable band structure and valley occupation in four-layer \mos through a combination of magnetotransport experiments and a self-consistent hybrid $k\cdot p$ tight-binding model.
At low gate voltages, we confirm that the conduction band minimum is centered at the $Q$ point, in line with theoretical expectations for multilayer \mos. 
However, as the gate voltage increases, charge redistribution towards the layer closest to the positive gate electrode induces a transition of the band edge from the $Q$ valleys to the $K$ valleys.
By extending our model to bilayer and three-layer \mos, we successfully bridge previous discrepancies between experimental observations \cite{pisoni_absence_2019, masseroni_electron_2021} and density functional theory predictions \cite{mattheiss_band_1973, li_electronic_2007, lebegue_electronic_2009, ellis_indirect_2011, yun_thickness_2012, sun_indirect--direct_2016}, underscoring the importance of interlayer screening and layer-specific charge accumulation in determining valley occupation.
These findings highlight the intricate interplay between the $K$ and $Q$ valleys in multilayer MoS$_2$, paving the way for future valleytronic and electronic applications that exploit layer-specific charge and valley control.

\section*{Acknowledgments}
We thank Peter Märki, Thomas Bähler, as well as the FIRST staff for their technical support.
We acknowledge support from the European Graphene Flagship Core 3 Project, Swiss National Science Foundation via NCCR Quantum Science, and H2020 European Research Council (ERC) Synergy Grant under Grant Agreement 95154. This work was supported by EPSRC Grants EP/S030719/1
and EP/V007033/1, and the Lloyd Register Foundation Nanotechnology Grant. Collaboration between ETH and the University of Manchester was supported by the International Science Partnerships Fund (UK).
I.S. and X.L. acknowledge financial support from the University of Manchester’s Dean’s Doctoral Scholarship. I.R. gratefully acknowledges the support of CARA Fellowship and BA/Cara/Leverhulme Research Support Grant LTRSF24/100044.
K.W. and T.T. acknowledge support from the JSPS KAKENHI (Grant Numbers 21H05233 and 23H02052) , the CREST (JPMJCR24A5), JST and World Premier International Research Center Initiative (WPI), MEXT, Japan.


\begin{thebibliography}{23}%
\makeatletter
\providecommand \@ifxundefined [1]{%
 \@ifx{#1\undefined}
}%
\providecommand \@ifnum [1]{%
 \ifnum #1\expandafter \@firstoftwo
 \else \expandafter \@secondoftwo
 \fi
}%
\providecommand \@ifx [1]{%
 \ifx #1\expandafter \@firstoftwo
 \else \expandafter \@secondoftwo
 \fi
}%
\providecommand \natexlab [1]{#1}%
\providecommand \enquote  [1]{``#1''}%
\providecommand \bibnamefont  [1]{#1}%
\providecommand \bibfnamefont [1]{#1}%
\providecommand \citenamefont [1]{#1}%
\providecommand \href@noop [0]{\@secondoftwo}%
\providecommand \href [0]{\begingroup \@sanitize@url \@href}%
\providecommand \@href[1]{\@@startlink{#1}\@@href}%
\providecommand \@@href[1]{\endgroup#1\@@endlink}%
\providecommand \@sanitize@url [0]{\catcode `\\12\catcode `\$12\catcode `\&12\catcode `\#12\catcode `\^12\catcode `\_12\catcode `\%12\relax}%
\providecommand \@@startlink[1]{}%
\providecommand \@@endlink[0]{}%
\providecommand \url  [0]{\begingroup\@sanitize@url \@url }%
\providecommand \@url [1]{\endgroup\@href {#1}{\urlprefix }}%
\providecommand \urlprefix  [0]{URL }%
\providecommand \Eprint [0]{\href }%
\providecommand \doibase [0]{https://doi.org/}%
\providecommand \selectlanguage [0]{\@gobble}%
\providecommand \bibinfo  [0]{\@secondoftwo}%
\providecommand \bibfield  [0]{\@secondoftwo}%
\providecommand \translation [1]{[#1]}%
\providecommand \BibitemOpen [0]{}%
\providecommand \bibitemStop [0]{}%
\providecommand \bibitemNoStop [0]{.\EOS\space}%
\providecommand \EOS [0]{\spacefactor3000\relax}%
\providecommand \BibitemShut  [1]{\csname bibitem#1\endcsname}%
\let\auto@bib@innerbib\@empty
\bibitem [{\citenamefont {Kim}\ \emph {et~al.}(2012)\citenamefont {Kim}, \citenamefont {Konar}, \citenamefont {Hwang}, \citenamefont {Lee}, \citenamefont {Lee}, \citenamefont {Yang}, \citenamefont {Jung}, \citenamefont {Kim}, \citenamefont {Yoo}, \citenamefont {Choi}, \citenamefont {Jin}, \citenamefont {Lee}, \citenamefont {Jena}, \citenamefont {Choi},\ and\ \citenamefont {Kim}}]{kim_high-mobility_2012}%
  \BibitemOpen
  \bibfield  {author} {\bibinfo {author} {\bibfnamefont {S.}~\bibnamefont {Kim}}, \bibinfo {author} {\bibfnamefont {A.}~\bibnamefont {Konar}}, \bibinfo {author} {\bibfnamefont {W.-S.}\ \bibnamefont {Hwang}}, \bibinfo {author} {\bibfnamefont {J.~H.}\ \bibnamefont {Lee}}, \bibinfo {author} {\bibfnamefont {J.}~\bibnamefont {Lee}}, \bibinfo {author} {\bibfnamefont {J.}~\bibnamefont {Yang}}, \bibinfo {author} {\bibfnamefont {C.}~\bibnamefont {Jung}}, \bibinfo {author} {\bibfnamefont {H.}~\bibnamefont {Kim}}, \bibinfo {author} {\bibfnamefont {J.-B.}\ \bibnamefont {Yoo}}, \bibinfo {author} {\bibfnamefont {J.-Y.}\ \bibnamefont {Choi}}, \bibinfo {author} {\bibfnamefont {Y.~W.}\ \bibnamefont {Jin}}, \bibinfo {author} {\bibfnamefont {S.~Y.}\ \bibnamefont {Lee}}, \bibinfo {author} {\bibfnamefont {D.}~\bibnamefont {Jena}}, \bibinfo {author} {\bibfnamefont {W.}~\bibnamefont {Choi}},\ and\ \bibinfo {author} {\bibfnamefont {K.}~\bibnamefont {Kim}},\ }\bibfield  {title} {\bibinfo {title} {High-mobility and low-power thin-film
  transistors based on multilayer {MoS}$_2$ crystals},\ }\href {https://doi.org/10.1038/ncomms2018} {\bibfield  {journal} {\bibinfo  {journal} {Nature Communications}\ }\textbf {\bibinfo {volume} {3}},\ \bibinfo {pages} {1011} (\bibinfo {year} {2012})}\BibitemShut {NoStop}%
\bibitem [{\citenamefont {Movva}\ \emph {et~al.}(2015)\citenamefont {Movva}, \citenamefont {Rai}, \citenamefont {Kang}, \citenamefont {Kim}, \citenamefont {Fallahazad}, \citenamefont {Taniguchi}, \citenamefont {Watanabe}, \citenamefont {Tutuc},\ and\ \citenamefont {Banerjee}}]{movva_high-mobility_2015}%
  \BibitemOpen
  \bibfield  {author} {\bibinfo {author} {\bibfnamefont {H.~C.~P.}\ \bibnamefont {Movva}}, \bibinfo {author} {\bibfnamefont {A.}~\bibnamefont {Rai}}, \bibinfo {author} {\bibfnamefont {S.}~\bibnamefont {Kang}}, \bibinfo {author} {\bibfnamefont {K.}~\bibnamefont {Kim}}, \bibinfo {author} {\bibfnamefont {B.}~\bibnamefont {Fallahazad}}, \bibinfo {author} {\bibfnamefont {T.}~\bibnamefont {Taniguchi}}, \bibinfo {author} {\bibfnamefont {K.}~\bibnamefont {Watanabe}}, \bibinfo {author} {\bibfnamefont {E.}~\bibnamefont {Tutuc}},\ and\ \bibinfo {author} {\bibfnamefont {S.~K.}\ \bibnamefont {Banerjee}},\ }\bibfield  {title} {\bibinfo {title} {High-{Mobility} {Holes} in {Dual}-{Gated} {WSe2} {Field}-{Effect} {Transistors}},\ }\href {https://doi.org/10.1021/acsnano.5b04611} {\bibfield  {journal} {\bibinfo  {journal} {ACS Nano}\ }\textbf {\bibinfo {volume} {9}},\ \bibinfo {pages} {10402} (\bibinfo {year} {2015})}\BibitemShut {NoStop}%
\bibitem [{\citenamefont {Sebastian}\ \emph {et~al.}(2021)\citenamefont {Sebastian}, \citenamefont {Pendurthi}, \citenamefont {Choudhury}, \citenamefont {Redwing},\ and\ \citenamefont {Das}}]{sebastian_benchmarking_2021}%
  \BibitemOpen
  \bibfield  {author} {\bibinfo {author} {\bibfnamefont {A.}~\bibnamefont {Sebastian}}, \bibinfo {author} {\bibfnamefont {R.}~\bibnamefont {Pendurthi}}, \bibinfo {author} {\bibfnamefont {T.~H.}\ \bibnamefont {Choudhury}}, \bibinfo {author} {\bibfnamefont {J.~M.}\ \bibnamefont {Redwing}},\ and\ \bibinfo {author} {\bibfnamefont {S.}~\bibnamefont {Das}},\ }\bibfield  {title} {\bibinfo {title} {Benchmarking monolayer {MoS}2 and {WS}2 field-effect transistors},\ }\href {https://doi.org/10.1038/s41467-020-20732-w} {\bibfield  {journal} {\bibinfo  {journal} {Nature Communications}\ }\textbf {\bibinfo {volume} {12}},\ \bibinfo {pages} {693} (\bibinfo {year} {2021})}\BibitemShut {NoStop}%
\bibitem [{\citenamefont {Wang}\ \emph {et~al.}(2012)\citenamefont {Wang}, \citenamefont {Kalantar-Zadeh}, \citenamefont {Kis}, \citenamefont {Coleman},\ and\ \citenamefont {Strano}}]{wang_electronics_2012}%
  \BibitemOpen
  \bibfield  {author} {\bibinfo {author} {\bibfnamefont {Q.~H.}\ \bibnamefont {Wang}}, \bibinfo {author} {\bibfnamefont {K.}~\bibnamefont {Kalantar-Zadeh}}, \bibinfo {author} {\bibfnamefont {A.}~\bibnamefont {Kis}}, \bibinfo {author} {\bibfnamefont {J.~N.}\ \bibnamefont {Coleman}},\ and\ \bibinfo {author} {\bibfnamefont {M.~S.}\ \bibnamefont {Strano}},\ }\bibfield  {title} {\bibinfo {title} {Electronics and optoelectronics of two-dimensional transition metal dichalcogenides},\ }\href {https://doi.org/10.1038/nnano.2012.193} {\bibfield  {journal} {\bibinfo  {journal} {Nature Nanotechnology}\ }\textbf {\bibinfo {volume} {7}},\ \bibinfo {pages} {699} (\bibinfo {year} {2012})}\BibitemShut {NoStop}%
\bibitem [{\citenamefont {Geim}\ and\ \citenamefont {Grigorieva}(2013)}]{geim_van_2013}%
  \BibitemOpen
  \bibfield  {author} {\bibinfo {author} {\bibfnamefont {A.~K.}\ \bibnamefont {Geim}}\ and\ \bibinfo {author} {\bibfnamefont {I.~V.}\ \bibnamefont {Grigorieva}},\ }\bibfield  {title} {\bibinfo {title} {Van der waals heterostructures},\ }\href {https://doi.org/10.1038/nature12385} {\bibfield  {journal} {\bibinfo  {journal} {Nature}\ }\textbf {\bibinfo {volume} {499}},\ \bibinfo {pages} {419} (\bibinfo {year} {2013})}\BibitemShut {NoStop}%
\bibitem [{\citenamefont {Kormányos}\ \emph {et~al.}(2015)\citenamefont {Kormányos}, \citenamefont {Burkard}, \citenamefont {Gmitra}, \citenamefont {Fabian}, \citenamefont {Zólyomi}, \citenamefont {Drummond},\ and\ \citenamefont {Fal'ko}}]{kormanyos_kp_theory_2015}%
  \BibitemOpen
  \bibfield  {author} {\bibinfo {author} {\bibfnamefont {A.}~\bibnamefont {Kormányos}}, \bibinfo {author} {\bibfnamefont {G.}~\bibnamefont {Burkard}}, \bibinfo {author} {\bibfnamefont {M.}~\bibnamefont {Gmitra}}, \bibinfo {author} {\bibfnamefont {J.}~\bibnamefont {Fabian}}, \bibinfo {author} {\bibfnamefont {V.}~\bibnamefont {Zólyomi}}, \bibinfo {author} {\bibfnamefont {N.~D.}\ \bibnamefont {Drummond}},\ and\ \bibinfo {author} {\bibfnamefont {V.}~\bibnamefont {Fal'ko}},\ }\bibfield  {title} {\bibinfo {title} {k$\cdot$p theory for two-dimensional transition metal dichalcogenide semiconductors},\ }\href {https://doi.org/10.1088/2053-1583/2/2/022001} {\bibfield  {journal} {\bibinfo  {journal} {2D Materials}\ }\textbf {\bibinfo {volume} {2}},\ \bibinfo {pages} {022001} (\bibinfo {year} {2015})}\BibitemShut {NoStop}%
\bibitem [{\citenamefont {Mak}\ \emph {et~al.}(2010)\citenamefont {Mak}, \citenamefont {Lee}, \citenamefont {Hone}, \citenamefont {Shan},\ and\ \citenamefont {Heinz}}]{mak_atomically_2010}%
  \BibitemOpen
  \bibfield  {author} {\bibinfo {author} {\bibfnamefont {K.~F.}\ \bibnamefont {Mak}}, \bibinfo {author} {\bibfnamefont {C.}~\bibnamefont {Lee}}, \bibinfo {author} {\bibfnamefont {J.}~\bibnamefont {Hone}}, \bibinfo {author} {\bibfnamefont {J.}~\bibnamefont {Shan}},\ and\ \bibinfo {author} {\bibfnamefont {T.~F.}\ \bibnamefont {Heinz}},\ }\bibfield  {title} {\bibinfo {title} {Atomically {Thin} $\mathrm{MoS_2}$: {A} {New} {Direct}-{Gap} {Semiconductor}},\ }\href {https://doi.org/10.1103/PhysRevLett.105.136805} {\bibfield  {journal} {\bibinfo  {journal} {Physical Review Letters}\ }\textbf {\bibinfo {volume} {105}},\ \bibinfo {pages} {136805} (\bibinfo {year} {2010})}\BibitemShut {NoStop}%
\bibitem [{\citenamefont {Splendiani}\ \emph {et~al.}(2010)\citenamefont {Splendiani}, \citenamefont {Sun}, \citenamefont {Zhang}, \citenamefont {Li}, \citenamefont {Kim}, \citenamefont {Chim}, \citenamefont {Galli},\ and\ \citenamefont {Wang}}]{splendiani_emerging_2010}%
  \BibitemOpen
  \bibfield  {author} {\bibinfo {author} {\bibfnamefont {A.}~\bibnamefont {Splendiani}}, \bibinfo {author} {\bibfnamefont {L.}~\bibnamefont {Sun}}, \bibinfo {author} {\bibfnamefont {Y.}~\bibnamefont {Zhang}}, \bibinfo {author} {\bibfnamefont {T.}~\bibnamefont {Li}}, \bibinfo {author} {\bibfnamefont {J.}~\bibnamefont {Kim}}, \bibinfo {author} {\bibfnamefont {C.-Y.}\ \bibnamefont {Chim}}, \bibinfo {author} {\bibfnamefont {G.}~\bibnamefont {Galli}},\ and\ \bibinfo {author} {\bibfnamefont {F.}~\bibnamefont {Wang}},\ }\bibfield  {title} {\bibinfo {title} {Emerging photoluminescence in monolayer {MoS} $_{\textrm{2}}$},\ }\href {https://doi.org/10.1021/nl903868w} {\bibfield  {journal} {\bibinfo  {journal} {Nano Letters}\ }\textbf {\bibinfo {volume} {10}},\ \bibinfo {pages} {1271} (\bibinfo {year} {2010})}\BibitemShut {NoStop}%
\bibitem [{\citenamefont {Pisoni}\ \emph {et~al.}(2018)\citenamefont {Pisoni}, \citenamefont {Kormányos}, \citenamefont {Brooks}, \citenamefont {Lei}, \citenamefont {Back}, \citenamefont {Eich}, \citenamefont {Overweg}, \citenamefont {Lee}, \citenamefont {Rickhaus}, \citenamefont {Watanabe}, \citenamefont {Taniguchi}, \citenamefont {Imamoglu}, \citenamefont {Burkard}, \citenamefont {Ihn},\ and\ \citenamefont {Ensslin}}]{pisoni_interactions_2018}%
  \BibitemOpen
  \bibfield  {author} {\bibinfo {author} {\bibfnamefont {R.}~\bibnamefont {Pisoni}}, \bibinfo {author} {\bibfnamefont {A.}~\bibnamefont {Kormányos}}, \bibinfo {author} {\bibfnamefont {M.}~\bibnamefont {Brooks}}, \bibinfo {author} {\bibfnamefont {Z.}~\bibnamefont {Lei}}, \bibinfo {author} {\bibfnamefont {P.}~\bibnamefont {Back}}, \bibinfo {author} {\bibfnamefont {M.}~\bibnamefont {Eich}}, \bibinfo {author} {\bibfnamefont {H.}~\bibnamefont {Overweg}}, \bibinfo {author} {\bibfnamefont {Y.}~\bibnamefont {Lee}}, \bibinfo {author} {\bibfnamefont {P.}~\bibnamefont {Rickhaus}}, \bibinfo {author} {\bibfnamefont {K.}~\bibnamefont {Watanabe}}, \bibinfo {author} {\bibfnamefont {T.}~\bibnamefont {Taniguchi}}, \bibinfo {author} {\bibfnamefont {A.}~\bibnamefont {Imamoglu}}, \bibinfo {author} {\bibfnamefont {G.}~\bibnamefont {Burkard}}, \bibinfo {author} {\bibfnamefont {T.}~\bibnamefont {Ihn}},\ and\ \bibinfo {author} {\bibfnamefont {K.}~\bibnamefont {Ensslin}},\ }\bibfield  {title} {\bibinfo {title} {Interactions and
  magnetotransport through spin-valley coupled landau levels in monolayer $\mathrm{MoS_2}$},\ }\href {https://doi.org/10.1103/PhysRevLett.121.247701} {\bibfield  {journal} {\bibinfo  {journal} {Physical Review Letters}\ }\textbf {\bibinfo {volume} {121}},\ \bibinfo {pages} {247701} (\bibinfo {year} {2018})}\BibitemShut {NoStop}%
\bibitem [{\citenamefont {Pisoni}\ \emph {et~al.}(2019)\citenamefont {Pisoni}, \citenamefont {Davatz}, \citenamefont {Watanabe}, \citenamefont {Taniguchi}, \citenamefont {Ihn},\ and\ \citenamefont {Ensslin}}]{pisoni_absence_2019}%
  \BibitemOpen
  \bibfield  {author} {\bibinfo {author} {\bibfnamefont {R.}~\bibnamefont {Pisoni}}, \bibinfo {author} {\bibfnamefont {T.}~\bibnamefont {Davatz}}, \bibinfo {author} {\bibfnamefont {K.}~\bibnamefont {Watanabe}}, \bibinfo {author} {\bibfnamefont {T.}~\bibnamefont {Taniguchi}}, \bibinfo {author} {\bibfnamefont {T.}~\bibnamefont {Ihn}},\ and\ \bibinfo {author} {\bibfnamefont {K.}~\bibnamefont {Ensslin}},\ }\bibfield  {title} {\bibinfo {title} {Absence of {Interlayer} {Tunnel} {Coupling} of ${K}$-{Valley} {Electrons} in {Bilayer} $\mathrm{MoS}_{2}$},\ }\href {https://doi.org/10.1103/PhysRevLett.123.117702} {\bibfield  {journal} {\bibinfo  {journal} {Physical Review Letters}\ }\textbf {\bibinfo {volume} {123}},\ \bibinfo {pages} {117702} (\bibinfo {year} {2019})}\BibitemShut {NoStop}%
\bibitem [{\citenamefont {Masseroni}\ \emph {et~al.}(2021)\citenamefont {Masseroni}, \citenamefont {Davatz}, \citenamefont {Pisoni}, \citenamefont {de~Vries}, \citenamefont {Rickhaus}, \citenamefont {Taniguchi}, \citenamefont {Watanabe}, \citenamefont {Fal'ko}, \citenamefont {Ihn},\ and\ \citenamefont {Ensslin}}]{masseroni_electron_2021}%
  \BibitemOpen
  \bibfield  {author} {\bibinfo {author} {\bibfnamefont {M.}~\bibnamefont {Masseroni}}, \bibinfo {author} {\bibfnamefont {T.}~\bibnamefont {Davatz}}, \bibinfo {author} {\bibfnamefont {R.}~\bibnamefont {Pisoni}}, \bibinfo {author} {\bibfnamefont {F.~K.}\ \bibnamefont {de~Vries}}, \bibinfo {author} {\bibfnamefont {P.}~\bibnamefont {Rickhaus}}, \bibinfo {author} {\bibfnamefont {T.}~\bibnamefont {Taniguchi}}, \bibinfo {author} {\bibfnamefont {K.}~\bibnamefont {Watanabe}}, \bibinfo {author} {\bibfnamefont {V.}~\bibnamefont {Fal'ko}}, \bibinfo {author} {\bibfnamefont {T.}~\bibnamefont {Ihn}},\ and\ \bibinfo {author} {\bibfnamefont {K.}~\bibnamefont {Ensslin}},\ }\bibfield  {title} {\bibinfo {title} {Electron transport in dual-gated three-layer $\mathrm{MoS}_{2}$},\ }\href {https://doi.org/10.1103/PhysRevResearch.3.023047} {\bibfield  {journal} {\bibinfo  {journal} {Physical Review Research}\ }\textbf {\bibinfo {volume} {3}},\ \bibinfo {pages} {023047} (\bibinfo {year} {2021})}\BibitemShut {NoStop}%
\bibitem [{\citenamefont {Mattheiss}(1973)}]{mattheiss_band_1973}%
  \BibitemOpen
  \bibfield  {author} {\bibinfo {author} {\bibfnamefont {L.~F.}\ \bibnamefont {Mattheiss}},\ }\bibfield  {title} {\bibinfo {title} {Band structures of transition-metal-dichalcogenide layer compounds},\ }\href {https://doi.org/10.1103/PhysRevB.8.3719} {\bibfield  {journal} {\bibinfo  {journal} {Physical Review B}\ }\textbf {\bibinfo {volume} {8}},\ \bibinfo {pages} {3719} (\bibinfo {year} {1973})}\BibitemShut {NoStop}%
\bibitem [{\citenamefont {Li}\ and\ \citenamefont {Galli}(2007)}]{li_electronic_2007}%
  \BibitemOpen
  \bibfield  {author} {\bibinfo {author} {\bibfnamefont {T.}~\bibnamefont {Li}}\ and\ \bibinfo {author} {\bibfnamefont {G.}~\bibnamefont {Galli}},\ }\bibfield  {title} {\bibinfo {title} {Electronic {Properties} of {MoS2} {Nanoparticles}},\ }\href {https://doi.org/10.1021/jp075424v} {\bibfield  {journal} {\bibinfo  {journal} {The Journal of Physical Chemistry C}\ }\textbf {\bibinfo {volume} {111}},\ \bibinfo {pages} {16192} (\bibinfo {year} {2007})}\BibitemShut {NoStop}%
\bibitem [{\citenamefont {Lebègue}\ and\ \citenamefont {Eriksson}(2009)}]{lebegue_electronic_2009}%
  \BibitemOpen
  \bibfield  {author} {\bibinfo {author} {\bibfnamefont {S.}~\bibnamefont {Lebègue}}\ and\ \bibinfo {author} {\bibfnamefont {O.}~\bibnamefont {Eriksson}},\ }\bibfield  {title} {\bibinfo {title} {Electronic structure of two-dimensional crystals from ab initio theory},\ }\href {https://doi.org/10.1103/PhysRevB.79.115409} {\bibfield  {journal} {\bibinfo  {journal} {Physical Review B}\ }\textbf {\bibinfo {volume} {79}},\ \bibinfo {pages} {115409} (\bibinfo {year} {2009})},\ \bibinfo {note} {publisher: American Physical Society}\BibitemShut {NoStop}%
\bibitem [{\citenamefont {Ellis}\ \emph {et~al.}(2011)\citenamefont {Ellis}, \citenamefont {Lucero},\ and\ \citenamefont {Scuseria}}]{ellis_indirect_2011}%
  \BibitemOpen
  \bibfield  {author} {\bibinfo {author} {\bibfnamefont {J.~K.}\ \bibnamefont {Ellis}}, \bibinfo {author} {\bibfnamefont {M.~J.}\ \bibnamefont {Lucero}},\ and\ \bibinfo {author} {\bibfnamefont {G.~E.}\ \bibnamefont {Scuseria}},\ }\bibfield  {title} {\bibinfo {title} {The indirect to direct band gap transition in multilayered {MoS2} as predicted by screened hybrid density functional theory},\ }\href {https://doi.org/10.1063/1.3672219} {\bibfield  {journal} {\bibinfo  {journal} {Applied Physics Letters}\ }\textbf {\bibinfo {volume} {99}},\ \bibinfo {pages} {261908} (\bibinfo {year} {2011})}\BibitemShut {NoStop}%
\bibitem [{\citenamefont {Yun}\ \emph {et~al.}(2012)\citenamefont {Yun}, \citenamefont {Han}, \citenamefont {Hong}, \citenamefont {Kim},\ and\ \citenamefont {Lee}}]{yun_thickness_2012}%
  \BibitemOpen
  \bibfield  {author} {\bibinfo {author} {\bibfnamefont {W.~S.}\ \bibnamefont {Yun}}, \bibinfo {author} {\bibfnamefont {S.~W.}\ \bibnamefont {Han}}, \bibinfo {author} {\bibfnamefont {S.~C.}\ \bibnamefont {Hong}}, \bibinfo {author} {\bibfnamefont {I.~G.}\ \bibnamefont {Kim}},\ and\ \bibinfo {author} {\bibfnamefont {J.~D.}\ \bibnamefont {Lee}},\ }\bibfield  {title} {\bibinfo {title} {Thickness and strain effects on electronic structures of transition metal dichalcogenides: 2{H}-{M}{X}$_2$ semiconductors ( {M} = {Mo}, {W}; {X} = {S}, {Se}, {Te})},\ }\href {https://doi.org/10.1103/PhysRevB.85.033305} {\bibfield  {journal} {\bibinfo  {journal} {Physical Review B}\ }\textbf {\bibinfo {volume} {85}},\ \bibinfo {pages} {033305} (\bibinfo {year} {2012})}\BibitemShut {NoStop}%
\bibitem [{\citenamefont {Sun}\ \emph {et~al.}(2016)\citenamefont {Sun}, \citenamefont {Wang},\ and\ \citenamefont {Shuai}}]{sun_indirect--direct_2016}%
  \BibitemOpen
  \bibfield  {author} {\bibinfo {author} {\bibfnamefont {Y.}~\bibnamefont {Sun}}, \bibinfo {author} {\bibfnamefont {D.}~\bibnamefont {Wang}},\ and\ \bibinfo {author} {\bibfnamefont {Z.}~\bibnamefont {Shuai}},\ }\bibfield  {title} {\bibinfo {title} {Indirect-to-direct band gap crossover in few-layer transition metal dichalcogenides: A theoretical prediction},\ }\href {https://doi.org/10.1021/acs.jpcc.6b08748} {\bibfield  {journal} {\bibinfo  {journal} {The Journal of Physical Chemistry C}\ }\textbf {\bibinfo {volume} {120}},\ \bibinfo {pages} {21866} (\bibinfo {year} {2016})}\BibitemShut {NoStop}%
\bibitem [{\citenamefont {Movva}\ \emph {et~al.}(2018)\citenamefont {Movva}, \citenamefont {Lovorn}, \citenamefont {Fallahazad}, \citenamefont {Larentis}, \citenamefont {Kim}, \citenamefont {Taniguchi}, \citenamefont {Watanabe}, \citenamefont {Banerjee}, \citenamefont {MacDonald},\ and\ \citenamefont {Tutuc}}]{movva_tunable_2018}%
  \BibitemOpen
  \bibfield  {author} {\bibinfo {author} {\bibfnamefont {H.~C.}\ \bibnamefont {Movva}}, \bibinfo {author} {\bibfnamefont {T.}~\bibnamefont {Lovorn}}, \bibinfo {author} {\bibfnamefont {B.}~\bibnamefont {Fallahazad}}, \bibinfo {author} {\bibfnamefont {S.}~\bibnamefont {Larentis}}, \bibinfo {author} {\bibfnamefont {K.}~\bibnamefont {Kim}}, \bibinfo {author} {\bibfnamefont {T.}~\bibnamefont {Taniguchi}}, \bibinfo {author} {\bibfnamefont {K.}~\bibnamefont {Watanabe}}, \bibinfo {author} {\bibfnamefont {S.~K.}\ \bibnamefont {Banerjee}}, \bibinfo {author} {\bibfnamefont {A.~H.}\ \bibnamefont {MacDonald}},\ and\ \bibinfo {author} {\bibfnamefont {E.}~\bibnamefont {Tutuc}},\ }\bibfield  {title} {\bibinfo {title} {Tunable ${\Gamma}$-${K}$ {Valley} {Populations} in {Hole}-{Doped} {Trilayer} $\mathrm{WSe_2}$},\ }\href {https://doi.org/10.1103/PhysRevLett.120.107703} {\bibfield  {journal} {\bibinfo  {journal} {Physical Review Letters}\ }\textbf {\bibinfo {volume} {120}},\ \bibinfo {pages} {107703} (\bibinfo {year}
  {2018})}\BibitemShut {NoStop}%
\bibitem [{\citenamefont {Fallahazad}\ \emph {et~al.}(2016)\citenamefont {Fallahazad}, \citenamefont {Movva}, \citenamefont {Kim}, \citenamefont {Larentis}, \citenamefont {Taniguchi}, \citenamefont {Watanabe}, \citenamefont {Banerjee},\ and\ \citenamefont {Tutuc}}]{fallahazad_shubnikovhaas_2016}%
  \BibitemOpen
  \bibfield  {author} {\bibinfo {author} {\bibfnamefont {B.}~\bibnamefont {Fallahazad}}, \bibinfo {author} {\bibfnamefont {H.~C.~P.}\ \bibnamefont {Movva}}, \bibinfo {author} {\bibfnamefont {K.}~\bibnamefont {Kim}}, \bibinfo {author} {\bibfnamefont {S.}~\bibnamefont {Larentis}}, \bibinfo {author} {\bibfnamefont {T.}~\bibnamefont {Taniguchi}}, \bibinfo {author} {\bibfnamefont {K.}~\bibnamefont {Watanabe}}, \bibinfo {author} {\bibfnamefont {S.~K.}\ \bibnamefont {Banerjee}},\ and\ \bibinfo {author} {\bibfnamefont {E.}~\bibnamefont {Tutuc}},\ }\bibfield  {title} {\bibinfo {title} {Shubnikov–de {Haas} {Oscillations} of {High}-{Mobility} {Holes} in {Monolayer} and {Bilayer} {WSe}$_2$ : {Landau} {Level} {Degeneracy}, {Effective} {Mass}, and {Negative} {Compressibility}},\ }\href {https://doi.org/10.1103/PhysRevLett.116.086601} {\bibfield  {journal} {\bibinfo  {journal} {Physical Review Letters}\ }\textbf {\bibinfo {volume} {116}},\ \bibinfo {pages} {086601} (\bibinfo {year} {2016})}\BibitemShut {NoStop}%
\bibitem [{\citenamefont {Larentis}\ \emph {et~al.}(2018)\citenamefont {Larentis}, \citenamefont {Movva}, \citenamefont {Fallahazad}, \citenamefont {Kim}, \citenamefont {Behroozi}, \citenamefont {Taniguchi}, \citenamefont {Watanabe}, \citenamefont {Banerjee},\ and\ \citenamefont {Tutuc}}]{larentis_large_2018}%
  \BibitemOpen
  \bibfield  {author} {\bibinfo {author} {\bibfnamefont {S.}~\bibnamefont {Larentis}}, \bibinfo {author} {\bibfnamefont {H.~C.~P.}\ \bibnamefont {Movva}}, \bibinfo {author} {\bibfnamefont {B.}~\bibnamefont {Fallahazad}}, \bibinfo {author} {\bibfnamefont {K.}~\bibnamefont {Kim}}, \bibinfo {author} {\bibfnamefont {A.}~\bibnamefont {Behroozi}}, \bibinfo {author} {\bibfnamefont {T.}~\bibnamefont {Taniguchi}}, \bibinfo {author} {\bibfnamefont {K.}~\bibnamefont {Watanabe}}, \bibinfo {author} {\bibfnamefont {S.~K.}\ \bibnamefont {Banerjee}},\ and\ \bibinfo {author} {\bibfnamefont {E.}~\bibnamefont {Tutuc}},\ }\bibfield  {title} {\bibinfo {title} {Large effective mass and interaction-enhanced {Zeeman} splitting of ${K}$-valley electrons in $\mathrm{MoSe_2}$},\ }\href {https://doi.org/10.1103/PhysRevB.97.201407} {\bibfield  {journal} {\bibinfo  {journal} {Physical Review B}\ }\textbf {\bibinfo {volume} {97}},\ \bibinfo {pages} {201407} (\bibinfo {year} {2018})}\BibitemShut {NoStop}%
\bibitem [{\citenamefont {Masseroni}\ \emph {et~al.}(2023)\citenamefont {Masseroni}, \citenamefont {Qu}, \citenamefont {Taniguchi}, \citenamefont {Watanabe}, \citenamefont {Ihn},\ and\ \citenamefont {Ensslin}}]{masseroni_evidence_2023}%
  \BibitemOpen
  \bibfield  {author} {\bibinfo {author} {\bibfnamefont {M.}~\bibnamefont {Masseroni}}, \bibinfo {author} {\bibfnamefont {T.}~\bibnamefont {Qu}}, \bibinfo {author} {\bibfnamefont {T.}~\bibnamefont {Taniguchi}}, \bibinfo {author} {\bibfnamefont {K.}~\bibnamefont {Watanabe}}, \bibinfo {author} {\bibfnamefont {T.}~\bibnamefont {Ihn}},\ and\ \bibinfo {author} {\bibfnamefont {K.}~\bibnamefont {Ensslin}},\ }\bibfield  {title} {\bibinfo {title} {Evidence of the coulomb gap in the density of states of ${\mathrm{mos}}_{2}$},\ }\href {https://doi.org/10.1103/PhysRevResearch.5.013113} {\bibfield  {journal} {\bibinfo  {journal} {Phys. Rev. Res.}\ }\textbf {\bibinfo {volume} {5}},\ \bibinfo {pages} {013113} (\bibinfo {year} {2023})}\BibitemShut {NoStop}%
\bibitem [{\citenamefont {Rozhansky}\ and\ \citenamefont {Fal'ko}(2024)}]{PhysRevB.110.L161404}%
  \BibitemOpen
  \bibfield  {author} {\bibinfo {author} {\bibfnamefont {I.}~\bibnamefont {Rozhansky}}\ and\ \bibinfo {author} {\bibfnamefont {V.}~\bibnamefont {Fal'ko}},\ }\bibfield  {title} {\bibinfo {title} {Exchange-enhanced spin-orbit splitting and its density dependence for electrons in monolayer transition metal dichalcogenides},\ }\href {https://doi.org/10.1103/PhysRevB.110.L161404} {\bibfield  {journal} {\bibinfo  {journal} {Phys. Rev. B}\ }\textbf {\bibinfo {volume} {110}},\ \bibinfo {pages} {L161404} (\bibinfo {year} {2024})}\BibitemShut {NoStop}%
\bibitem [{\citenamefont {Rickhaus}\ \emph {et~al.}(2019)\citenamefont {Rickhaus}, \citenamefont {Zheng}, \citenamefont {Lado}, \citenamefont {Lee}, \citenamefont {Kurzmann}, \citenamefont {Eich}, \citenamefont {Pisoni}, \citenamefont {Tong}, \citenamefont {Garreis}, \citenamefont {Gold}, \citenamefont {Masseroni}, \citenamefont {Taniguchi}, \citenamefont {Wantanabe}, \citenamefont {Ihn},\ and\ \citenamefont {Ensslin}}]{rickhaus_gap_2019}%
  \BibitemOpen
  \bibfield  {author} {\bibinfo {author} {\bibfnamefont {P.}~\bibnamefont {Rickhaus}}, \bibinfo {author} {\bibfnamefont {G.}~\bibnamefont {Zheng}}, \bibinfo {author} {\bibfnamefont {J.~L.}\ \bibnamefont {Lado}}, \bibinfo {author} {\bibfnamefont {Y.}~\bibnamefont {Lee}}, \bibinfo {author} {\bibfnamefont {A.}~\bibnamefont {Kurzmann}}, \bibinfo {author} {\bibfnamefont {M.}~\bibnamefont {Eich}}, \bibinfo {author} {\bibfnamefont {R.}~\bibnamefont {Pisoni}}, \bibinfo {author} {\bibfnamefont {C.}~\bibnamefont {Tong}}, \bibinfo {author} {\bibfnamefont {R.}~\bibnamefont {Garreis}}, \bibinfo {author} {\bibfnamefont {C.}~\bibnamefont {Gold}}, \bibinfo {author} {\bibfnamefont {M.}~\bibnamefont {Masseroni}}, \bibinfo {author} {\bibfnamefont {T.}~\bibnamefont {Taniguchi}}, \bibinfo {author} {\bibfnamefont {K.}~\bibnamefont {Wantanabe}}, \bibinfo {author} {\bibfnamefont {T.}~\bibnamefont {Ihn}},\ and\ \bibinfo {author} {\bibfnamefont {K.}~\bibnamefont {Ensslin}},\ }\bibfield  {title} {\bibinfo {title} {Gap {Opening} in
  {Twisted} {Double} {Bilayer} {Graphene} by {Crystal} {Fields}},\ }\href {https://doi.org/10.1021/acs.nanolett.9b03660} {\bibfield  {journal} {\bibinfo  {journal} {Nano Letters}\ }\textbf {\bibinfo {volume} {19}},\ \bibinfo {pages} {8821} (\bibinfo {year} {2019})}\BibitemShut {NoStop}%
\end{thebibliography}
%

\section*{Author contributions}
M.M., T.I., K.E. conceived and designed the experiments.
M.M. fabricated the device with inputs from T.I. and K.E..
M.M., performed the measurements with inputs from T.I and K.E..
M.M. and A.S. analyzed the experimental data with inputs from M.N., T.I., and K.E..
I.S. and V.F. developed the theoretical model with inputs from J.G.M., I.R., and X.L..
M.M. designed the figures with inputs from I.S..
T.T., K.W. supplied the hexagonal boron nitride.
M.M wrote the manuscript with inputs from I.S., J.G.M. and I.R..
All the coauthors mentioned above read and commented on the manuscript.
M.M. and I.S. contributed equally to this work.

\end{document}